\documentclass[aps,prl,a4paper,twocolumn,superscriptaddress,showkeys,longbibliography]{revtex4-1}

\usepackage[colorlinks=true,bookmarks=false,citecolor=blue,linkcolor=red,hyperfootnotes=true,urlcolor=blue]{hyperref}

\usepackage{physics}
\usepackage{graphicx}
\usepackage{dcolumn}
\usepackage{bm}
\usepackage[dvipsnames]{xcolor}
\usepackage{wasysym}
\usepackage{lipsum}
\usepackage{enumitem}
\usepackage{booktabs} 
\usepackage{gensymb}
\usepackage{xcolor}
\usepackage{amssymb}
\usepackage{booktabs}
\usepackage{array}
\newcolumntype{L}{>{$}l<{$}}
\newcolumntype{C}{>{$}c<{$}} 

\usepackage{amsmath}

\newcolumntype{R}{>{$}r<{$}}

\usepackage{mathtools}
\usepackage{siunitx}
\usepackage{soul}
\usepackage{float}

\makeatletter
\def\p@subsection{}
\makeatother

\begin{document}

\title{Bipolaronic superconductivity out of a Coulomb gas}

\author{J. Sous} \thanks{These authors contributed equally: J. Sous. C. Zhang}
\affiliation{Department of Physics, Stanford University, Stanford, CA 93405, USA}
\affiliation{Stanford Institute for Theoretical Physics, Stanford University, Stanford, CA5, USA}

\author{C. Zhang}\thanks{These authors contributed equally: C. Zhang, J. Sous.} \thanks{Author to whom correspondence should be addressed.} \email{czhang@lps.ecnu.edu.cn}
\affiliation{State Key Laboratory of Precision Spectroscopy,
East China Normal University, Shanghai 200062, China}

\author{M. Berciu} 
\affiliation{Department of Physics and Astronomy, University of British Columbia, Vancouver, British Columbia V6T 1Z1, Canada}
\affiliation{Stewart Blusson Quantum Matter Institute, University of British Columbia, Vancouver, British Columbia V6T 1Z4, Canada}

\author{D. R. Reichman} 
\affiliation{Department of Chemistry, Columbia University, New York,
New York 10027, USA}

\author{B. V. Svistunov}
\affiliation{Department of Physics, University of Massachusetts, Amherst, Massachusetts 01003, USA}
\affiliation{Wilczek Quantum Center, School of Physics and Astronomy and T. D. Lee Institute, Shanghai Jiao Tong University, Shanghai 200240, China}

\author{N. V. Prokof’ev}
\affiliation{Department of Physics, University of Massachusetts, Amherst, Massachusetts 01003, USA}

\author{A. J. Millis} \thanks{Author to whom correspondence should be addressed} \email{ajm2010@columbia.edu}
\affiliation{Department of Physics, Columbia University, New York, New York 10027, USA} 
\affiliation{Center for Computational Quantum Physics, Flatiron Institute, 162 5$^{th}$ Avenue, New York, NY 10010, USA}

\date{\today}

\begin{abstract}
Employing unbiased sign-problem-free quantum Monte Carlo, we investigate the effects of long-range Coulomb forces on BEC of bipolarons using a model of  bond phonon-modulated electron hopping. In absence of long-range repulsion, this model was recently shown to give rise to small-size, light-mass bipolarons that undergo a superfluid transition at high values of the critical transition temperature $T_\mathrm{c}$.  We find that $T_\mathrm{c}$ in our model even with the long-range Coulomb repulsion remains much larger than that of Holstein bipolarons, and can be on the order of or greater than  the typical upper bounds on phonon-mediated $T_\mathrm{c}$ based on the Migdal-Eliashberg and McMillan approximations. Our work points to a physically simple mechanism for superconductivity in the low-density regime that may be relevant to current experiments on dilute superconductors.
\end{abstract}

\maketitle

\emph{Introduction.}
Understanding the mechanisms of superconductivity in the dilute density regime is an active theme of research, relevant to polar materials~\cite{STO,STOReview}, doped topological insulators~\cite{DopedTI,DopedTI2},  transition metal dichalcogenides~\cite{TMDSC}, moir\'e materials~\cite{TBG1,TBG2,TBG3,TTG1,TTG2,TTG3} and other materials~\cite{Others}. This large and growing list of ultra-low carrier density superconductors motivates theoretical examination of superconductivity in electron-phonon coupled systems at very small densities where, as a matter of principle, the Fermi liquid/Migdal-Eliashberg paradigm must fail. Bose-Einstein condensation (BEC) of preformed pairs (``bipolarons'') in principle offers a robust route to superconductivity at low densities. But, in the low-density regime, the Coulomb repulsion is weakly screened and thus the pairing ``glue'' required to bind electron pairs into bound states must be strong enough to overcome the Coulomb repulsion.  A strong pairing interaction is  usually believed to result in heavy bound states, implying low values of the critical transition temperature $T_\mathrm{c}$. These considerations~\cite{CRFBipolaron,KivelsonGeballe} are widely believed to severely limit the maximum $T_\mathrm{c}$ obtained from phonon-mediated binding of electrons into bipolarons.

We have recently shown~\cite{BipolaronicHighTc} that even in the presence of a short-ranged interaction parameterized by a large onsite Hubbard repulsion $U$, electrons coupled to phonons via bond phonon-modulated electron hopping form small-size, light-mass bipolarons~\cite{SousBipolaron,QMCBondBipolaron} that undergo a superfluid (``BEC'') transition at values of $T_\mathrm{c}$ that are much larger than those obtained in (Holstein) models in which the electron density is coupled to phonons  or from  Migdal-Eliashberg theory of superconductivity out of a Fermi liquid. This work did not include the long-range part of the Coulomb interaction, so is relevant to two-dimensional (2D) materials in which the Coulomb repulsion is completely screened by gating or proximity to a substrate~\cite{BipolaronicHighTc}. However, in ungated 2D materials and in three-dimensional (3D) materials in which the Coulomb repulsion cannot be screened by an external gate, the question of the effects of long-range Coulomb repulsion on bipolaronic superconductivity (and other non-phononic BEC mechanisms~\cite{HighTcReview, LFu1,LFu2,LFu3,LFu4}) remains open.

In this letter, we study BEC of bipolarons occuring in a dilute, 3D Coulomb gas, showing that $T_\mathrm{c}$ of bond-coupled bipolarons is still higher than that of density-coupled (Holstein) bipolarons, and in line with the value of $T_\mathrm{c}$ typically found in experiments on 3D materials believed to be close to or in the low-density regime.  In a Coulomb system, the two-electron bound state retains a finite size even at the critical interaction strength associated with unbinding~\cite{2bodycriticality}, and thus the maximum $T_\mathrm{c}$ is determined by a combination of binding strength, mass and size with the constraint that the size cannot be infinite. To the best of our knowledge, despite decades of debate~\cite{Alexandrov2000,Pacha1,CRFBipolaron}, our theory is the first quantitative effort that takes the presence of long-range Coulomb interaction into account and (i) demonstrates, using an unbiased approach, a realistic mechanism for BEC formation at relatively high values of $T_\mathrm{c}$ and (ii) unveils the properties of bipolarons, e.g. their mass and size, in 3D.

\begin{figure}[!]
\raggedright
\includegraphics[width=0.92\columnwidth]{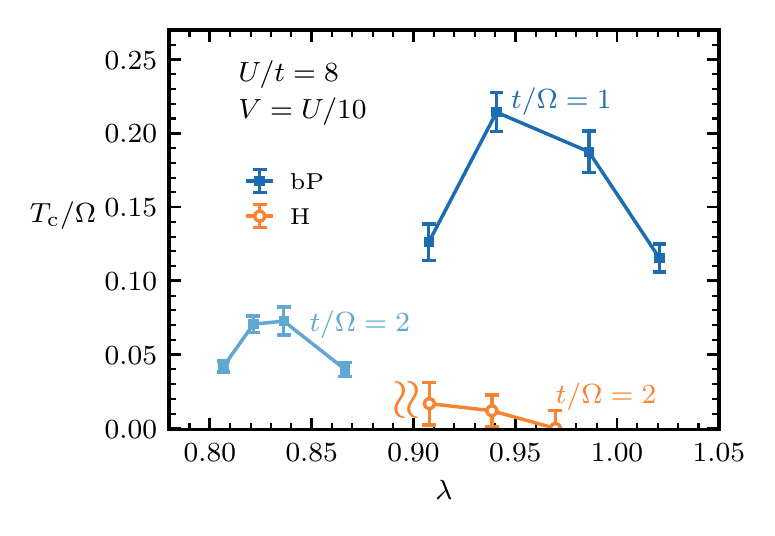}
\vspace{-4mm}
\caption{Estimates of $T_\mathrm{c}$ of the bond-Peierls (bP) bipolaronic superconductor (filled squares, blue lines) for different $t/\Omega$  at $U = 8t$, ${\mathcal{V}}(r>0)=V/r$ with $V=U/10$  as a function of $\lambda$ computed according to Eq.~\eqref{Eq:Tc} from QMC simulations of the bipolaron effective mass $m^{\! \star}_\mathrm{BP}$ and its mean squared-radius $R^2_{\mathrm{BP}}$. We contrast this to superconductivity of Holstein (H) bipolarons (open circles, orange line) at $t/\Omega = 2$ for the same values of $U/t$ and $V$. Here we use $\lambda = \alpha^2/3\Omega t$ for bond-Peierls bipolarons, $\lambda = 0.85 \alpha^2/6\Omega t$ for Holstein bipolarons (we use a factor of $0.85$ so that the two sets of data can be presented on the same scale of $\lambda$). The doubly wavy symbol indicates absence of bipolarons in the Holstein model for $\lambda \lesssim 0.91$ where a crossover from BEC to BCS may occur.}
\vspace{-4mm}
\label{fig:Fig1}
\end{figure}

\begin{figure}[!t]
\raggedright
\vspace{-3mm}
\includegraphics[width=0.92\columnwidth]{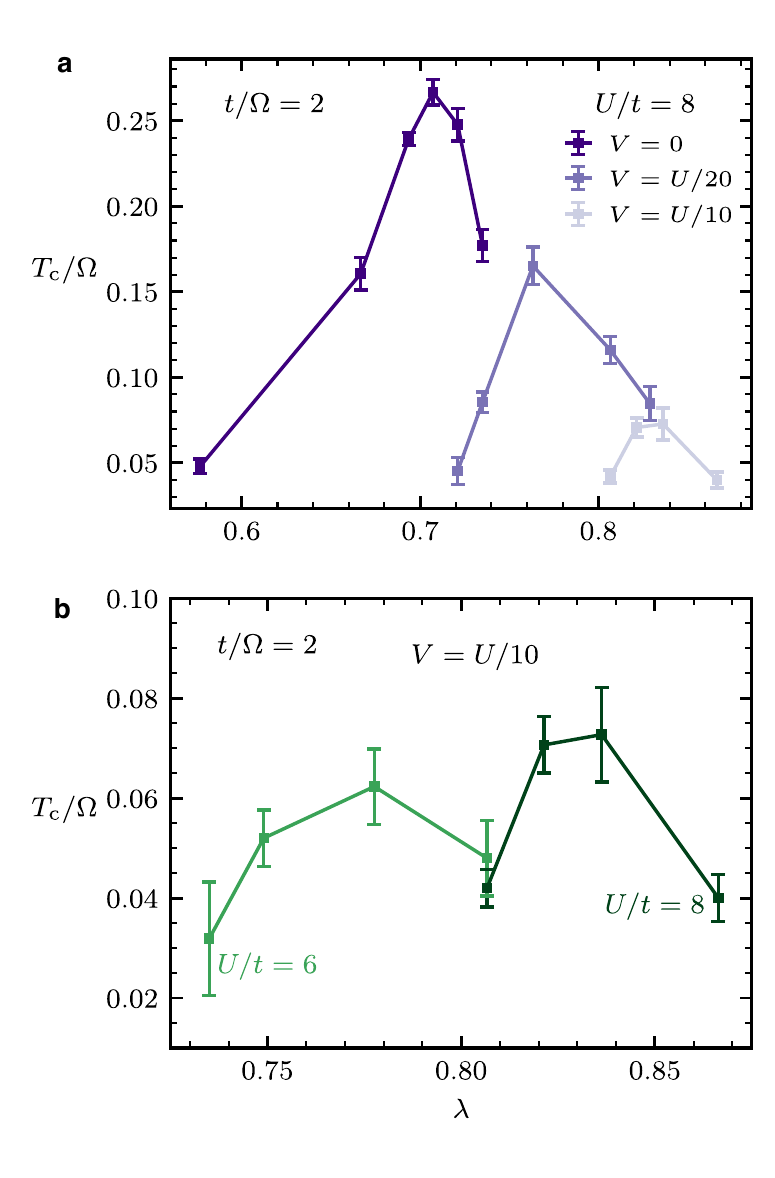}
\vspace{-6mm}
\caption{Estimates of $T_\mathrm{c}$ of the bond-Peierls bipolaronic superconductor for  $t/\Omega = 2$ as a function of $\lambda = \alpha^2/(3\Omega t)$, for {\bf a.} different strengths of the Coulomb repulsion $V$ at fixed onsite $U/t=8$, and for {\bf b.} different strengths of the onsite $U$ at fixed long-range repulsion $V=U/10$, computed according to Eq.~\eqref{Eq:Tc} from QMC simulations of the bipolaron effective mass $m^{\! \star}_\mathrm{BP}$ and mean squared-radius $R^2_{\mathrm{BP}}$.}
\label{fig:Fig2}
\vspace{-4mm}     
\end{figure}

\begin{figure*}
\centering
\includegraphics[width=1.778\columnwidth]{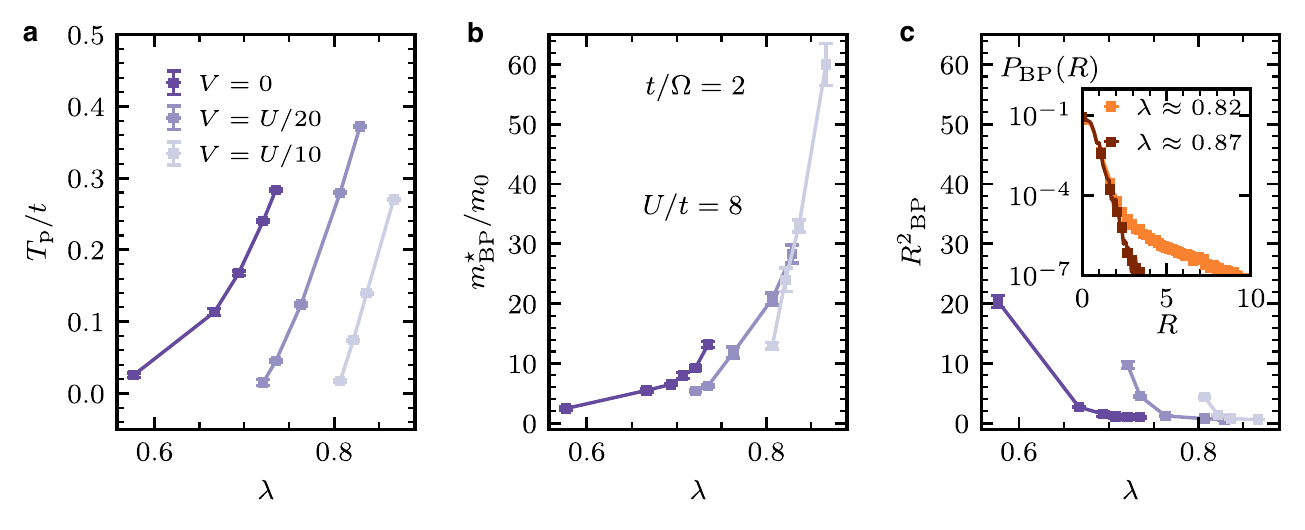}
\vspace{-4mm}
\caption{Bipolaron properties in the model, Eq.~\eqref{Eq:H}, at $t/\Omega = 2$ as a function of  $\lambda = \alpha^2/(3\Omega t)$ for different $V$ at fixed $U/t=8$: {\bf a}. Bipolaron binding energy $T_{\mathrm{p}}$, {\bf b}. Bipolaron effective mass $m^{\! \star}_\mathrm{BP}$ in units of the mass of two free electrons $m_0 = 2m_e = 1/t$,  and {\bf c}. Bipolaron squared-radius $R^2_{\mathrm{BP}}$ and, in the inset, its radial size probability density distribution $P_\mathrm{BP}(R)$ at large coupling. Error bars in $P_\mathrm{BP}(R)$ correspond to statistical errors smaller than the symbol size and therefore are not shown.}\label{fig:Fig3}
\vspace{-4mm}
\end{figure*}

\emph{Model.} We consider the bond-Peierls~\cite{ BarisicPeierls1} (also known as bond-Su-Schrieffer-Heeger~\cite{su1979solitons}) electron-phonon coupling on a 3D cubic lattice. In this model the electronic hopping between two sites is modulated by a single oscillator centered on the bond connecting the two sites~\cite{BipolaronicHighTc}. The  Hamiltonian is
\begin{equation}
\hat{\cal H}=\hat{\cal H}_\mathrm{e} + \hat{\cal H}_\mathrm{ph} + \hat{\cal{V}}_\mathrm{{e\mbox{-}ph}}.
\label{Eq:H}
\end{equation}
Here, the lattice Coulomb model for electrons with spin $\sigma \in \{\uparrow,\downarrow\}$ is
$\hat{\cal H}_\mathrm{ e} =-t \sum_{\langle i,j \rangle, \sigma}^{}\left( \hat{c}_{i,\sigma}^\dagger \hat{c}_{j,\sigma} + \mathrm{h.c.}\right) +U \sum_{i}^{} \hat{n}_{i,\uparrow} \hat{n}_{i,\downarrow} +(1/2) \sum_{i\neq j}^{} \mathcal{V}_{i,j}\hat{n}_{i} \hat{n}_{j}$
with nearest-neighbor (NN) hopping $t$, onsite Hubbard repulsion $ U$, NN repulsion $V$ and longer-range repulsion $\mathcal{V}_{ij} = \frac{V a}{|r_i - r_j|}$, where $\hat n_{i} = \hat n_{i,\uparrow} + \hat n_{i,\downarrow}$ with $\hat n_{i,\sigma} = \hat{c}_{i,\sigma}^\dagger \hat{c}_{i,\sigma}$ at site $r_i$, and $a$ is the lattice constant (NN distance). The notation $\langle i,j \rangle$ refers to NN sites. Writing the Coulomb repulsion as $e^2/\epsilon r$ (where the dielectric constant $\epsilon$ encodes short-ranged polarization effects arising from degrees of freedom not explicitly included in the model) we estimate the onsite $U = e^2/\epsilon a_\mathrm{orb}$ with $a_\mathrm{orb}$ the unit cell orbital size, the NN $V = e^2/\epsilon a \approx U a_\mathrm{orb}/a$, and further-neighbor (distance $r>a$)  ${\mathcal{V}_{r>a}} = Va/r$.  We henceforth set the lattice constant $a=1$.  We model  distortions of the bonds connecting sites $i$ and $j$ as Einstein oscillators with Hamiltonian $\hat{\cal H}_\mathrm{ph} = \sum_{\langle i,j \rangle} \big(\frac{1}{2} K \hat{X}_{i,j}^2 + \hat{P}_{i,j}^2/2M \big) = \Omega \sum_{\langle i,j \rangle} \hat{b}_{i,j}^\dagger \hat{b}_{i,j}$ with oscillator frequency ($\hbar=1$) $\Omega = \sqrt{K/M}$. The interaction between electrons and phonons takes the form
\begin{eqnarray}
\hat{\cal{V}}_\mathrm{{e\mbox{-}ph}} &=& \alpha \sqrt{2M\Omega}  \sum_{\langle i,j \rangle,\sigma}^{}\left( \hat{c}_{i,\sigma}^\dagger \hat{c}_{j,\sigma} +
\mathrm{h.c.}\right) \hat{X}_{i,j}
\label{Eq:Veph}
\end{eqnarray}
describing the modulation of electron hopping by an oscillator $\hat{X}_{i,j} \coloneqq \frac{1}{\sqrt{2M\Omega}}  \left( \hat{b}_{i,j}^\dagger+ \hat{b}_{i,j} \right)$ associated with the bond connecting sites $i$ and $j$  with coupling coefficient $\alpha{\sqrt{2M\Omega}}$. We set $M=1$. The relevant parameters are a dimensionless coupling constant $\lambda = ((\alpha \sqrt{2\Omega})^2/K)/6t = \alpha^2/(3\Omega t)$, the ratio of the typical polaronic energy scale to the free electron energy scale, and an adiabaticity parameter $t/\Omega$.  It is important to note that  a typical physical origin for this phonon-modulated hopping is from  interference of different hopping pathways~\cite{BipolaronicHighTc}. This implies that the magnitude of the coupling term (Eq.~\eqref{Eq:Veph}) is independent of the bare hopping, which means that the model remains valid in the strong-coupling regime even when the electronic hopping changes sign. This is different from other models of phonon-modulated hopping~\cite{Capone1,Capone2,marchand2010sharp,SousBipolaron,ChaoPeierlsPolaron,CarbonePeierlsPolaron} in which
an equation of the general form of Eq.~\eqref{Eq:Veph} applies only in the small displacement regime where the net change in hopping amplitude is small relative to the bare hopping~\cite{QMCXrepresentation}.

\emph{Method.} Using a numerically exact sign-problem-free quantum Monte Carlo (QMC) approach based on a path-integral formulation of the electronic sector combined with either a real-space diagrammatic or a Fock-space path-integral representation of the phononic sector~\cite{QMCBondBipolaron} we study singlet bipolaron formation in the two-electron sector of the model.  We simulate the model on a 3D cubic lattice with linear size $L=140$ sites and open boundary conditions. This system size is large enough to access the thermodynamic limit and eliminate boundary effects. We expect that our results are qualitatively similar for other models in which the hopping is modulated by phonons.

\emph{BEC of bipolarons.} In 3D, a dilute system of electrons, at strong enough electron-phonon coupling, is unstable to the formation of bipolarons and thus forms a gas of interacting bosons.  Competing instabilities such as phase separation~\cite{NoceraPS} or Wigner crystallization~\cite{SpivakKivelson} are very unlikely in three dimensions in a Coulomb system unless the density is extremely low.  Thus, based on recent results on BEC out of a Coulomb Bose plasma~\cite{BosePlasma}, bipolarons will undergo condensation into a BEC and superfluidity at a $T \leq T_\mathrm{c}$ with $T_\mathrm{c} \approx 3.2 \rho_\mathrm{BP}^{2/3}/m^{\! \star}_\mathrm{BP}$, where  $\rho_\mathrm{BP}$ is the density of bipolarons and $m^{\!\star}_\mathrm{BP} \coloneqq [({\partial^2 E_{\mathrm{BP}} (K)}/{\partial K^2})|_{K=0}]^{-1}$ is the bipolaron effective mass, with $E_{\mathrm{BP}}(K)$ the bipolaron dispersion and $K$ the bipolaron momentum.  This estimate remains reliable in a broad density range up to the density at which bipolarons overlap.  The largest $T_\mathrm{c}$ from this mechanism
is thus expected to arise around the density $\rho_\mathrm{BP} = \min\{(1/(\frac{4}{3} \pi R^3_\mathrm{BP}), 1/(\frac{4}{3}\pi)\}$ at which the inter-particle separation becomes on the order of the bipolaron radial size $R_\mathrm{BP}$, which, after lattice regularization,  must be at least unity.  This leads to an estimate of $T_\mathrm{c}$ of the bipolaron superconductor at the overlap density  that depends only on the bipolaron properties given by
\begin{eqnarray}
T_\mathrm{c} \approx
    \begin{cases}
        \frac{1.2}{m^{\! \star}_\mathrm{BP} R^2_\mathrm{BP} } & \text{if $R^2_\mathrm{BP}\geq 1$}\\
        \frac{1.2}{m^{\! \star}_\mathrm{BP}} & \text{otherwise},
    \end{cases}
\label{Eq:Tc}    
\end{eqnarray}
where $R^2_{\mathrm{BP}}\coloneqq \bra{\Psi_{\mathrm{BP}}} \hat{R}^2 \ket{\Psi_{\mathrm{BP}}}$ is the bipolaron mean squared-radius, with $\Psi_{\mathrm{BP}}$ the bipolaron ground state wavefunction.

\paragraph{Bipolaronic superconductivity out of a Coulomb gas.}  Figure~\ref{fig:Fig1} presents $T_\mathrm{c}/\Omega$ computed from Eq.~\eqref{Eq:Tc} using $m^{\! \star}_\mathrm{BP}$ and $R^2_\mathrm{BP}$ obtained from QMC simulations of two fermions in the model as a function of $\lambda$ for different $t/\Omega$ at $U=8t$ and $V=U/10$. Error bars in the figures represent one standard deviation statistical errors in QMC simulations. The results in Fig.~\ref{fig:Fig1} show that a sufficiently large value of $\lambda$ is needed in 3D in order form bound states and obtain an $s$-wave bipolaronic superconductor in the strong correlation regime considered here. Note that, first, the choice of $U/t = 8$ implies strong competition between the onsite repulsion and the electronic kinetic energy, and, second, the ratio  $V/U = 1/10$ is a reasonable estimate for typical materials such as the transition metal oxides, which have a roughly  10:1 ratio between the lattice constant ($\sim 4$\AA) and the orbital size ($\sim 0.5$\AA). Thus, our results in Fig.~\ref{fig:Fig1} prove that  bipolaronic superconductivity is robust even in the presence of strong, poorly screened Coulomb repulsion, see Fig.~\ref{fig:Fig2} for more extensive results. To appreciate this result, we contrast in Fig.~\ref{fig:Fig1} $T_\mathrm{c} /\Omega$ to that of Holstein bipolarons~\cite{BoncaHBipolaron,MacridinBipolaron} with electron-phonon coupling term given by $\hat{\mathcal{V}}_\mathrm{{e\mbox{-}ph}} = \alpha \sqrt{2\Omega} \sum_{i,\sigma}^{}\hat{c}_{i,\sigma}^\dagger \hat{c}_{i,\sigma} \hat{X}_i$. We find that $T_\mathrm{c}$ of the bond-Peierls bipolaronic superconductor is generically higher than that in the Holstein model including in the adiabatic limit $t>\Omega$. $T_\mathrm{c}$ of the Holstein bipolaronic superconductor never exceeds $\sim 0.03\Omega$ because binding requires sufficiently large $\lambda$ which leads to either large size at criticality or rapid mass enhancement soon after and thus a low $T_\mathrm{c}$. In contrast, the bond-Peierls bipolaron becomes strongly bound but remains relatively light and small and therefore can accommodate the Coulomb repulsion, see Fig.~\ref{fig:Fig3}. We note that the values of $T_{\mathrm{c}}/\Omega$ in this case appear to be comparable to or greater than the upper bound of $0.05$ from McMillan's phenomenological approach to Migdal-Eliashberg theory in the adiabatic limit for moderate values of $\lambda \lesssim 1$~\footnote{McMillan formula ~\cite{McMillanFormula}: $\frac{T_\mathrm{c}}{\Omega} = \frac{1}{1.45} e^{  -1.04 \frac{1+\lambda}{\lambda-\mu^\star(1 + 0.62\lambda)}}$ (where $\mu^\star=0.12$ is the value of the Coulomb pseudopotential found in many materials) predicts a typical upper bound of $T_\mathrm{c}/\Omega \sim 0.05$ at $\lambda = 1$.}. In other words, our theory predicts values of $T_\mathrm{c}$ in line with typical values found in experiments on dilute superconductors.

\begin{figure}[!t]
\vspace{-2mm}  
\raggedright
\includegraphics[width=0.92\columnwidth]{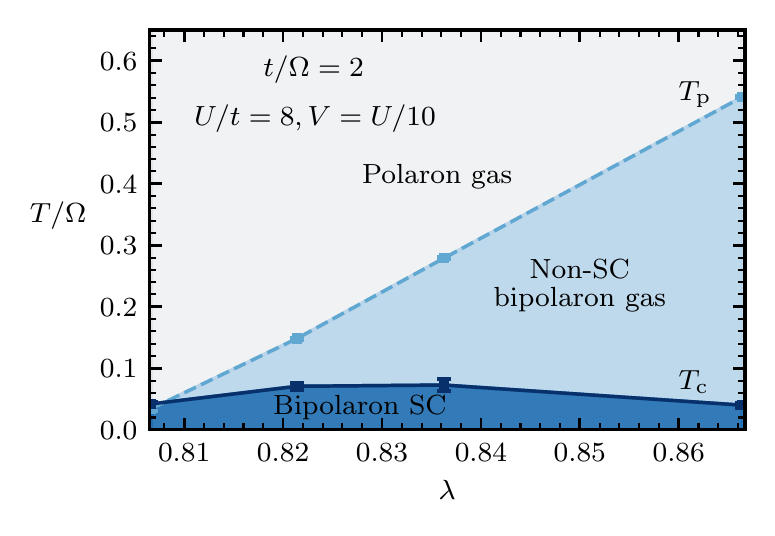}
\vspace{-4mm}  
\caption{Phase diagram in the $T/\Omega$-$\lambda$ space of the bond-Peierls model in the adiabatic limit $t/\Omega = 2$ with strong onsite $U/t=8$ and Coulomb $V=U/10$ interactions.  A BEC superconductor forms at $T\leq T_\mathrm{c}$ (dark blue region).  There is a large region at temperatures $T_\mathrm{c}<T \leq T_\mathrm{p}$ characterized by non-superconducting correlations and phase fluctuations in a normal gas of bipolarons  (light blue region).  Above $T_\mathrm{p}$, the bipolarons unbind into a polaron gas (gray region).}
\label{fig:Fig4}
\vspace{-4mm}     
\end{figure}

Having established bipolaronic superconductivity out of a Coulomb gas, we study in Fig.~\ref{fig:Fig2} the dependence of  $T_\mathrm{c}/\Omega$ on $V/U$ at fixed $U/t$ (Fig.~\ref{fig:Fig2}{\bf a}) and on $U/t$ at fixed $V/U$ (Fig.~\ref{fig:Fig2}{\bf b}).  The main results of Fig.~\ref{fig:Fig2}  can be summarized as follows. First, we find long-range repulsion-induced reduction of $T_\mathrm{c}$, but, importantly, $T_\mathrm{c}/\Omega$ remains relatively large even for large $V=U/10$.  This reflects the ability of the bipolaron wavefunction to spread itself effectively over multiple sites in order to accommodate the Coulomb repulsion, see inset of Fig.~\ref{fig:Fig3}{\bf c}. Second, onsite repulsion may enhance $T_\mathrm{c}$ similar to what we found in 2D models with no long-range interaction~\cite{BipolaronicHighTc}.  This unconventional behavior follows from the ability of the bipolaron to effectively reduce its mass without significantly increasing its radius as $U$ is increased. Our analysis thus shows that in presence of Coulomb repulsion bond bipolarons form with relatively light mass yet with small size and can benefit from the local repulsion by decreasing their mass without significantly increasing their size, hence proving that the  bond bipolaronic superconductivity is generally much less sensitive to Coulomb repulsion than Holstein bipolarons, and can in fact take advantage of the local repulsion to increase $T_\mathrm{c}$.

\begin{figure}[!b]
\vspace{-6mm}  
\raggedright
\includegraphics[width=\columnwidth]{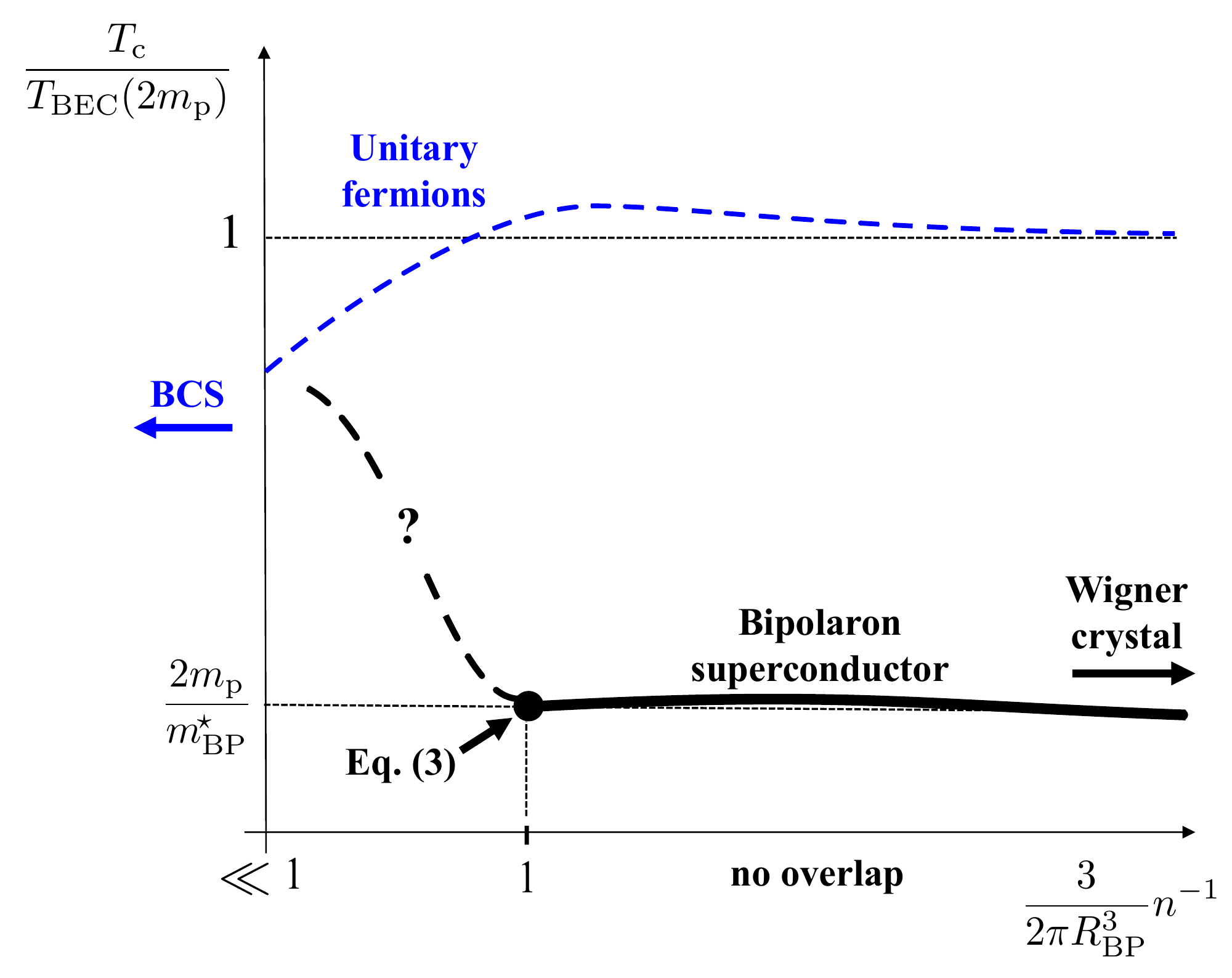}
\vspace{-8mm} 
\caption{Heavy black line: Superfluid transition temperature $T_\mathrm{c}$ of the bipolaronic gas  normalized to the BEC temperature $T_\mathrm{BEC}$ of a gas of bosons of density $n$ and mass $2m_\mathrm{p}$ as a function of the inverse  density $n^{-1}$ normalized to the effective bipolaron density $3/(2\pi R_\mathrm{BP}^3)$.   Dashed back line: Proposed extrapolation of $T_\mathrm{c}$ beyond the density at which the bipolarons overlap (black dot) and the BEC picture breaks down. As the density enters the $n>3/(2\pi R_\mathrm{BP}^3)$ BCS regime we expect, based on comparison to the unitary Fermi gas (dashed blue line), that the decrease in mass associated with unbinding of bipolarons into Fermi liquid quasiparticles may drive further increase in $T_\mathrm{c}$.}
\label{fig:Fig5}
\vspace{-4mm}     
\end{figure}


\paragraph{Phenomenology of bipolaronic superconductivity.}  Figure~\ref{fig:Fig3} details the features of bipolarons in presence of long-range Coulomb repulsion. A much sharper dependence on the electron-phonon coupling becomes apparent for larger values of $V$ as can be seen in the pairing temperature $T_\mathrm{p}$ (Fig.~\ref{fig:Fig3}{\bf a}) defined as the bipolaron binding energy and in the bipolaron effective mass (Fig.~\ref{fig:Fig3}{\bf b}) and squared-radius (Fig.~\ref{fig:Fig3}{\bf c}). This behavior implies that breakdown of Fermi liquid theory due to bipolaronic collapse depends non-trivially on the interplay of electron-phonon and electron-electron interactions and goes beyond the current understanding of the breakdown of Migdal-Eliasberg theory due to a first-order transition driven by bipolaron formation~\cite{ChakravertyBipolaron,ME_CDW1,ME_CDW2,MEbreakdownAlexandrov,CohenBounds,MEBreakdownMillis,MEBreakdownKivelson,TcBoundKivelson,GeneralHolsteinBipolaron}.  Second, Fig.~\ref{fig:Fig3} demonstrates that a sufficiently large value of $\lambda$ is needed in order to overcome the Coulomb repulsion and form bipolarons, but that mass enhancement  (Fig.~\ref{fig:Fig3}{\bf b}) remains moderate (mass enhancement is much more dramatic in the Holstein case).  We can also see from Fig.~\ref{fig:Fig3}{\bf b}, {\bf c} that small-size bipolarons with $R_{\mathrm{BP}} < 3$ exhibit rather weak mass enhancement $m^{\! \star}_\mathrm{BP} < 10$ explaining the large values of $T_\mathrm{c}$ at optimized $\lambda$ despite the strong Coulomb repulsion.  Overall, $T_\mathrm{c}$ drops from large values at $V=0$ to values approximately 3 times smaller at $V=U/10$, see also Fig.~\ref{fig:Fig2}.  This modest reduction in $T_\mathrm{c}$ reflects the ability of bipolarons to effectively accommodate the Coulomb repulsion by spreading their wavefunction effectively over multiple sites even at large $\lambda$, see the inset of Fig.~\ref{fig:Fig3}{\bf c}.  In Fig.~\ref{fig:Fig4} we study the phase diagram of the bipolaronic superconductor in the $T$-$\lambda$ space in the adiabatic limit in presence of large Coulomb repulsion.  A bipolaronic BEC superconductor forms at $T\leq T_\mathrm{c}$.  There is an extended region at temperatures $T_\mathrm{c}<T \leq T_\mathrm{p}$ characterized by non-superconducting correlations in a normal gas of bipolarons, which will give rise to a large regime of phase fluctuations in experiment.  Above $T_\mathrm{p}$, the bipolarons unbind into a polaron gas. Finally, in Fig.~\ref{fig:Fig5} we discuss the crossover or phase transition occurring as the density is increased into  the BCS regime~\cite{ModelQuarterFilling,fRGHubbardSSH} in which  bipolarons overlap. A useful analogy is provided by the unitary Fermi gas (fermions in the continuum interacting with attractive contact interaction) picture of the BCS-BEC crossover~\cite{UnitaryFermi1,UnitaryFermi2} where in the dilute limit weakly bound fermion pairs with mass twice the bare fermion mass condense into a BEC and as the density is increased into the BCS regime of overlapping  pairs $T_\mathrm{c}$ drops only slowly.  We may expect that the basic physics of the BCS-BEC crossover in the bipolaron system is similar to that for unitary fermions, but with the additional feature that the mass drops rapidly as the density increases beyond the point at which bipolarons overlap and unbind into polarons (with mass $m_\mathrm{p}$) and then into moderately renormalized electrons.  This density dependence of the mass, not present in the unitary Fermi gas, may lead to a further increase in $T_\mathrm{c}$ as the density is increased into the BCS regime.

\emph{Conclusion.}  We have shown using an exact approach that Bose condensation of bond-coupled bipolarons out of a Coulomb gas provides a robust mechanism for superconductivity in the dilute limit, with a transition temperature that is generically larger than what is found in the Holstein model even in the presence of strong Coulomb repulsion (as estimated for the  transition metal oxides for example).  This transition temperature can be greater than or on the order of the heuristic bounds on phonon-mediated superconductivity based on McMillan theory, suggesting that our mechanism may already be operative in a number of materials. We show that the local onsite Coulomb repulsion enhances $T_\mathrm{c}$ while the nonlocal part reduces $T_\mathrm{c}$, which, however, remains relatively high. The key ingredient of the combination of light mass and relatively small size of bipolarons even in presence of long-range Coulomb repulsion explains the robustness of the mechanism and the relatively large values of $T_\mathrm{c}$ found.  The crossover or phase transition to the BCS regime when bipolarons start to overlap is an open question, but, in contrast to the unitary Fermi gas~\cite{UnitaryFermi1,UnitaryFermi2}, we cannot rule out that $T_\mathrm{c}$ may have a peak as a function of carrier density in the crossover region where  bipolarons unbind into quasiparticles with much smaller mass which would imply even larger values of $T_\mathrm{c}$ on the BCS side.

\begin{acknowledgements}
We acknowledge useful discussions with S. Kivelson and S. Raghu. J.~S. acknowledges support from the Gordon and Betty Moore Foundation’s EPiQS Initiative through Grant GBMF8686 at Stanford University. J.~S., D.~R.~R. and A.~J.~M. acknowledge support from the National Science Foundation (NSF) Materials Research Science and Engineering Centers (MRSEC) program through Columbia University in the Center for Precision Assembly of Superstratic and Superatomic Solids under Grant No. DMR-1420634. C.~Z. acknowledges support from the National Natural Science Foundation of China (NSFC) under Grant No. 12204173.  M.~B. acknowledges support from the Natural Sciences and Engineering Research Council of Canada (NSERC), the Stewart Blusson Quantum Matter Institute (SBQMI) and the Max-Planck-UBC-UTokyo Center for Quantum Materials.  B.~V.~S. and N.~V.~P acknowledge support from the NSF under Grant No. DMR-2032077.  J.~S. also acknowledges the hospitality of the Center for Computational Quantum Physics (CCQ) at the Flatiron Institute. The Flatiron Institute is a division of the Simons Foundation.
\end{acknowledgements}

\catcode`'=9
\catcode``=9

\providecommand{\noopsort}[1]{}\providecommand{\singleletter}[1]{#1}%

\end{document}